\documentclass[apjl]{emulateapj}

\usepackage{hyperref}
\usepackage{amsmath}
\usepackage{amssymb}
\usepackage{graphicx}
\usepackage{color}
\usepackage{natbib}
\usepackage[normalem]{ulem}
\usepackage{multibib}

\def\be{\begin{equation}}
\def\ee{\end{equation}}
\def\ba{\begin{eqnarray}}
\def\ea{\end{eqnarray}}

\begin{document}

\title{An accurate analytical fit to the gravitational-wave inspiral duration for eccentric binaries}
\author{Ilya Mandel\altaffilmark{1,2,3}}
\email{ilya.mandel@monash.edu}
\affil{$^1$Monash Centre for Astrophysics, School of Physics and Astronomy, Monash University, Clayton, Victoria 3800, Australia}
\affil{$^2$The ARC Center of Excellence for Gravitational Wave Discovery -- OzGrav}
\affil{$^3$Institute of Gravitational Wave Astronomy and School of Physics and Astronomy, University of Birmingham, Birmingham, B15 2TT, United Kingdom}

\begin{abstract}
I present an analytical fit to the duration of the inspiral of two point masses driven by gravitational-wave emission.  The fit is accurate to within 3\% over the entire range of initial eccentricities from 0 to 0.99999.
\end{abstract}

\maketitle

\citet{Peters:1964} derived a very widely used post-Newtonian expression for the time $T$ for two point masses $M_1$ and $M_2$ to spiral in through gravitational-wave emission from an initial separation $a_0$ and initial eccentricity $e_0$:

\ba
T &=& T_c\, \frac{48}{19} \left(\frac{1-e_0^2}{e_0^{12/19} \left(1+\frac{121}{304}e_0^2\right)^{870/2299}}\right)^4 \nonumber\\
&\times& \int_0^{e_0} \frac{e^{29/19} \left(1+\frac{121}{304}e^2\right)^{1181/2299}}{(1-e^2)^{3/2} } de \label{eq:Peters},
\ea
where 
\be
T_c = \frac{5 c^5 a_0^4}{256 G^3 M_1 M_2 (M_1+M_2)}.
\ee

Equation (\ref{eq:Peters}) is somewhat awkward to compute in practice because it has apparently singular terms at both $e \to 0$ and $e \to 1$, although the complete expression is well-behaved in both limits.  Consequently it is common to use two approximations derived by \citet{Peters:1964}:
\be\label{eq:Tsmall}
\tau (e_0 \to 0) \approx T_c \left(\frac{1-e_0^2}{\left(1+\frac{121}{304}e_0^2\right)^{870/2299}}\right)^4
\ee
and
\be\label{eq:Tlarge}
\tau (e_0 \to 1) \approx \frac{768}{425} T_c \left(1-e_0^2\right)^{7/2}.
\ee

\begin{figure}
\centering
\includegraphics[width=\columnwidth]{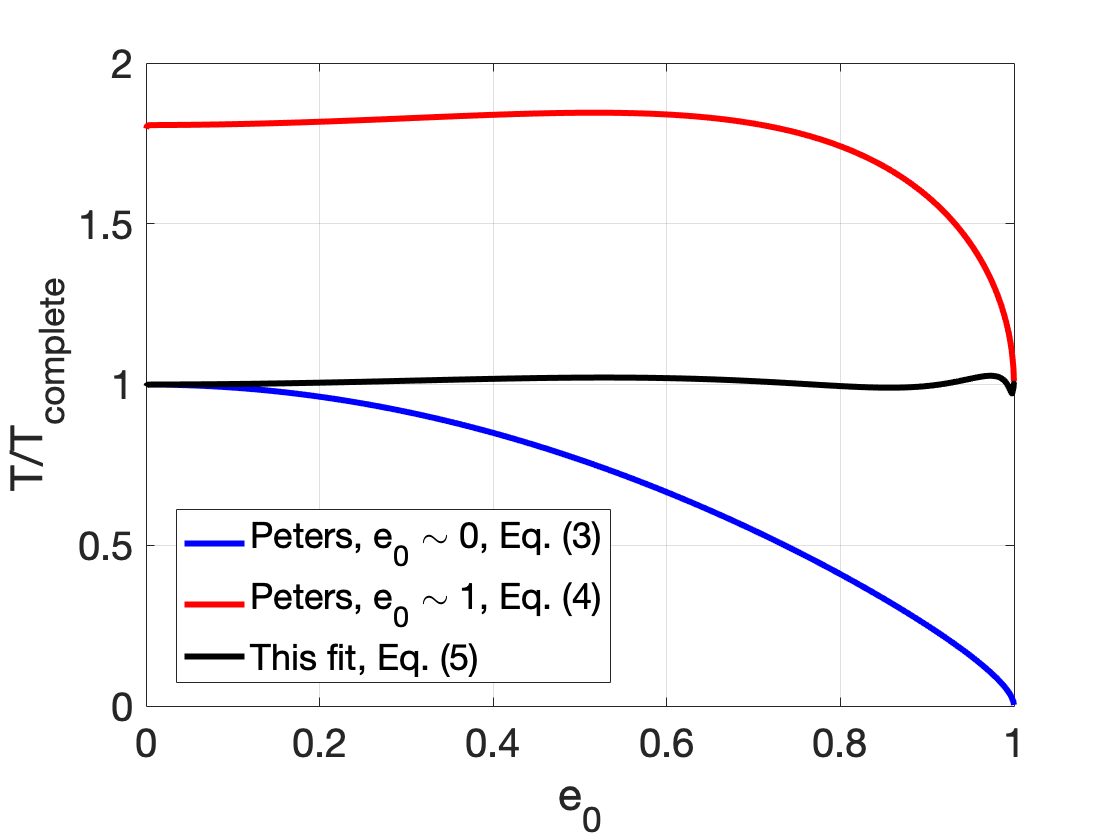}
\caption{The ratios of the fits given by Eqs.~(\ref{eq:Tsmall}), (\ref{eq:Tlarge}) and (\ref{eq:fit}) to the complete expression of Eq.~(\ref{eq:Peters}) as a function of initial eccentricity.  The fit proposed here (black) is accurate to within 3\% over the full range of initial eccentricities up to 0.99999.}\label{figure}
\end{figure}

However, as shown in Figure \ref{figure}, neither fit is particularly accurate in the intervening regime.  For example, at $e_0=0.8$, the estimate from Eq.~(\ref{eq:Tsmall}) is too low by a factor of 2.4, while that from Eq.~(\ref{eq:Tlarge}) is too large by 75\%.  Therefore, although sometimes one of the estimates (\ref{eq:Tsmall}) or (\ref{eq:Tlarge}) is used for all eccentricities, this is quite inaccurate.  Some codes that need to compute inspiral durations more accurately (e.g., binary population synthesis codes, \citealt{COMPAS:2021}) instead opt to split up the inspiral duration calculation into three pieces: the low-eccentricity regime where fit (\ref{eq:Tsmall}) is applied, the high-eccentricity regime where fit (\ref{eq:Tlarge}) is applied, and the intermediate regime where the integral (\ref{eq:Peters}) is computed numerically.  This, however, becomes rather cumbersome.  

\citet{PierroPinto:1996} showed that Eq.~(\ref{eq:Peters}) can be solved exactly with an Appell function, a 2-variable hypergeometric series.  However, to remain accurate, the series expansion requires a larger number of terms as eccentricity grows, and is not readily available in many software libraries.  Meanwhile, interpolation between pre-computed values of the inspiral duration can be an accurate and computationally efficient solution; however, the extra implementation effort makes it somewhat inconvenient.

The following fit to Eq.~(\ref{eq:Peters}) is purely analytical and easy to compute:
\be\label{eq:fit}
T \approx T_c \left(1+0.27 e_0^{10} + 0.33 e_0^{20} + 0.2 e_0^{1000}\right) (1-e_0^2)^{7/2}.
\ee
It is accurate to $<3\%$ over the entire range of initial eccentricities between 0 and 0.99999, i.e., $(1-e_0) \in [10^{-5},1]$.  The ratio of this fit to the full solution given by Eq.~(\ref{eq:Peters}) is shown in  Figure \ref{figure} in black.

The \citet{Peters:1964} solution given in Eq.~(\ref{eq:Peters}) is an approximation valid in the inspiral-dominated post-Newtonian regime, which is generally the situation of interest for gravitational-wave sources with non-extreme mass ratios that are captured with periapsis separations significantly larger than the innermost stable circular orbit (see \citealt{PierroPinto:1996} for a detailed discussion of the regime of applicability).  A number of works relaxed one or more of these conditions to obtain more complex expressions for the inspiral duration.  For example, \citet{Gair:2006time} computed corrections to the quadrupole flux from conservative relativistic dynamics in order to obtain durations of eccentric extreme-mass-ratio inspirals (EMRIs). \citet{WillMaitra:2017} included the effect of the spin of the supermassive black hole on the duration of an eccentric EMRI.  \citet{Zwick:2020} considered the impact of post-Newtonian corrections; however, their fit is less accurate at moderate eccentricities and non-monotonic as a function of eccentricity.  \citet{TuckerWill:2021} introduced a fit to the inspiral duration including post-Newtonian corrections through the 4.5 order, which is accurate to within 2\% for $100\leq a_0 (1-e_0^2) c^2 / G / (M_1+M_2) \leq 1000$ and $0 \leq e_0 \leq 0.999$.  

Nevertheless, the simple fit to the \citet{Peters:1964} gravitational-wave driven coalescence time proposed in Eq.~(\ref{eq:fit}) may prove useful in astrophysically relevant settings. 

\acknowledgements
I thank Bence Kocsis, Jonathan Gair, Cole Miller, Karthik Rajeev, Frank Timmes, Alexandria Tucker, Clifford Will, Xingjiang Zhu, and Lorenz Zwick for discussions. I acknowledge support from the Australian Research Council Centre of Excellence for Gravitational  Wave  Discovery  (OzGrav), through project number CE17010004. I am a recipient of the Australian Research Council Future Fellowship FT190100574.

\bibliographystyle{hapj}
\bibliography{Mandel}

\end{document}